\documentclass{emulateapj}
\usepackage{apjfonts}
\usepackage{amsmath}

\def\mr#1{\mathrm{#1}}

\newcounter{ichi}
\newcounter{ni}
\slugcomment{}

\shorttitle{High Energy Neutrinos and CRs from LL-GRBs?}
\shortauthors{Murase et al.}

\begin{document}


\title{High Energy Neutrinos and Cosmic-Rays \\
    from Low-Luminosity Gamma-Ray Bursts?}


\author{Kohta Murase\altaffilmark{1}, Kunihito Ioka\altaffilmark{2},
Shigehiro Nagataki\altaffilmark{1,3}, and Takashi Nakamura\altaffilmark{2}}


\altaffiltext{1}{YITP, Kyoto University, Kyoto, 606-8502, Japan}
\altaffiltext{2}{Department of Physics, Kyoto University, Kyoto 606-8502, Japan}
\altaffiltext{3}{KIPAC. Stanford University, Stanford, CA, 94309, USA}


\begin{abstract}
The recently discovered gamma-ray burst (GRB) GRB 060218/SN 2006aj
 is classified as an X-ray Flash with very long duration driven possibly
by a neutron star. Since GRB 060218 is very near $\sim 140$ Mpc and very dim,
one-year observation by \textit{Swift} suggests that the rate of GRB 060218-like events might be very high
so that such low luminosity GRBs (LL-GRBs) might form 
a different population from the cosmological high 
luminosity GRBs (HL-GRBs). We found that the high energy neutrino
 background from LL-GRBs could be comparable with that
 from HL-GRBs. If each neutrino event is detected by IceCube, 
later optical-infrared follow-up observations
such as by Subaru and HST have possibilities to identify a Type \Roman{ichi}bc
supernova associated with LL-GRBs, even if gamma- and X-rays 
are not observed by \textit{Swift}. This is in a sense a new
window from neutrino astronomy, which might enable us
to confirm the existence of LL-GRBs and to obtain information about
 their rate and origin. We also argue LL-GRBs as high energy gamma-ray 
and cosmic-ray sources.
\end{abstract}





\section{Introduction}
Gamma-ray bursts (GRBs) and supernovae (SNe) are most powerful
phenomena in the universe. Theorists predicted that the former would
result from the death of massive stars, and the association of
long-duration GRBs with core-collapse supernovae (SNe of Type
\Roman{ichi}bc to be more specific) has been observed over the last
decade. The first hint for such a connection came
with the discovery of a nearby SN 1998bw (SN \Roman{ichi}c) in the
error circle of GRB 980425 \cite{Gal1,Iwa1}. The first spectroscopic 
identification of a SN Ic superposed on a
GRB afterglow component was done in GRB 030329/SN 2003dh \cite{Hjo1}. 

Recently \textit{Swift} discovered GRB 060218, which is the second
nearest GRB identified to-date (Campana et al. 2006; Cusumano et
al. 2006; Mirabal \& Halpern 2006; Sakamoto 2006). 
GRB 060218 is associated with SN 2006aj and
provides another example of LL-GRBs. This event is 100 times
less energetic and the duration is very long $\sim 2000$ s. A thermal 
component in the X-ray and UV-optical spectra was discovered
and the size of the emitting black-body region is estimated to be
$r_{\mr{BB}}\sim (5 \times {10}^{11}-{10}^{12})$ cm. It would come
from the shock break out from a stellar envelope or a
dense wind \cite{Cam1}, or from a hot
cocoon surrounding the GRB ejecta \cite{Lia2,Ram1}.

These GRB 060218-like
events are phenomenologically peculiar events compared with the
conventional bursts because these have lower isotropic luminosity and
energy, simpler prompt light curves, and larger spectral time lags. 
Guetta et al. (2004)
argued that no bright burst $z<0.17$ should be observed by a
HETE-like instruments within the next 20 years, assuming that GRB
060218-like bursts follow the logN-logP relationship of HL-GRBs. 
Therefore, this unexpected discovery of GRB 060218 can lead
to the idea that they form a different new class of GRBs 
from the conventional HL-GRBs, although much uncertainty
remains and we do not know what distinguishes such LL-GRBs from the 
conventional HL-GRBs. Under this assumption, Soderberg et
al. (2006b) estimated the rate of LL-GRBs and found that they are about
ten times more common than conventional HL-GRBs, and Liang et
al. (2006a) obtained a high LL-GRB rate similarly. However, each origin of
LL-GRBs could be different. Mazzali et al. (2006) suggested that GRB
980425 and GRB 031203 could be related with a black hole formation,
while GRB 060218 could be driven by a neutron star. GRB 060218-like
events might possibly be associated with the birth of magnetars. The
origin of such LL-GRBs and whether these bursts are typical or not are
open problems. 

If such LL-GRBs like GRB 060218 are more common than HL-GRBs, they
would provide enough energy for high energy cosmic-rays, neutrinos,
and gamma-rays. In this Letter, we study the possibility of the high
energy cosmic-ray production and the successive neutrino production in LL-GRBs 
under the usual internal shock model.\footnote{When we were completing
the draft, we knew that a similar study was independently carried out by Gupta \& Zhang (2006).} 
Such neutrino bursts from
HL-GRBs were predicted in the context of the standard scenario
of GRBs assuming that ultra-high-energy cosmic rays (UHECRs) come from GRBs 
\cite{Wax1,Wax2}. Murase \& Nagataki (2006a, 2006b) also investigated 
such emission from HL-GRBs and from flares, using the Monte Carlo
simulation kit GEANT4 \cite{Ago1} with experimental data. With the
same method, we study the high energy neutrino emission from
LL-GRBs. We will also discuss various implications for
such LL-GRBs. Large neutrino detectors such as IceCube
\cite{Ahr1}, ANTARES \cite{Asl1} and NESTOR \cite{Gri1} are being
constructed. In the near future, these detectors may detect high
energy neutrino signals and give us more clues to understanding
LL-GRBs and testing our model.

\section{THE MODEL}
We suppose GRB 060218-like events as LL-GRBs in this Letter. 
GRB 060218 has low luminosity $\sim 10^{46-47}$ ergs/s, which is much 
smaller than that of usual HL-GRBs, typically  $L_{\mr{max}} \sim
{10}^{51-52}$ ergs/s. Hereafter, we take  $L_{\mr{max}} = 
{10}^{47} \, \mr{ergs/s}$ as a peak luminosity of LL-GRBs and fix
$E_{\gamma,\mr{iso}}=L_{\mr{max}} \delta t N \sim {10}^{49-50}$ ergs as
the released radiation energy. Here $\delta t$ is the variability time
and $N$ is the number of collisions.
To explain the prompt emission, we assume the usual internal shock model in
which the gamma-rays arise from the internal dissipation of 
ultra-relativistic jets, although there is another explanation \cite{Dai1}.
The typical collision radius will be expressed by commonly used relation, 
$r \approx {10}^{15}{(\Gamma/10)}^2 {(\delta t/150 \, \mr{s})}$ cm.
Of course, this radius has to be smaller than the deceleration radius, 
$r<r_{\mr{BM}} \approx 4.4 \times {10}^{16}
{(E_{\mr{kin},50}/n_{0} {(\Gamma/10)}^2)}^{1/3}$ cm. 
The observed light curve of GRB 060218 is simple and smooth, 
suggesting $\delta t \sim {10}^{2-3}$ s \cite{Cus1}. But it is
uncertain whether these parameters are typical or not \cite{Fan1}. 
Hence, we take 
$r \sim {10}^{14-16}$ cm with $\Gamma \sim (10-100)$.
These radii will be important for neutrino production \cite{KM1}. 
We also assume
that the Lorentz factor of the internal shocks will be mildly
relativistic, $\Gamma_\mr{sh} \approx
(\sqrt{\Gamma_{\mr{f}}/\Gamma_{\mr{s}}}+\sqrt{\Gamma_{\mr{s}}/\Gamma_{\mr{f}}})
/2 \sim$ a few.
The typical values in the usual synchrotron model are obtained as
follows. The minimum Lorentz
factor of electrons is estimated by $\gamma
_{e,\mr{m}} \approx \epsilon _e (m_p/m_e)(\Gamma _{\mr{sh}}-1)$.
Since the intensity of magnetic field is given by 
$B = 7.3
\times {10}^{2} \, \mr{G} \epsilon _{B,-1}^{1/2} {(\Gamma _{\mr{sh}}
(\Gamma _{\mr{sh}}-1)/2)}^{1/2} L _{\mr{M},48}^{1/2}
{(\Gamma/10)}^{-1} r_{15}^{-1}$,
the observed break energy is,
$E ^{\mr{b}}= \hbar \gamma_{e,\mr{m}}^2 \Gamma eB/m_e c
\sim 1 \, \, \mr{keV} \epsilon _{e}^2 \epsilon _{B}^{1/2} 
{(\Gamma _{\mr{sh}}-1)}^{5/2} {(\Gamma _{\mr{sh}}/2)}^{1/2} 
L _{\mr{M},48}^{1/2} r_{15}^{-1}$,
where $L_{\mr{M}}$ is the outflow luminosity. This value 
is not so different from the observed peak energy of GRB 060218, $
E^{\mr{b}} \sim$ keV. 

Although we have too less information about spectral
features of LL-GRBs at present, we assume a similar
spectral shape to that of HL-GRBs for our calculations and 
approximate it by the broken power-law instead of
exploiting a Band spectrum. The photon spectrum in the comoving frame
is expressed by, 
$dn/d\varepsilon = n_\mr{b}{(\varepsilon/
{\varepsilon}^{\mr{b}})}^{-\alpha}$ for $\varepsilon^{\mr{min}} 
< \varepsilon < \varepsilon ^{\mr{b}}$ and $dn/d\varepsilon= n_\mr{b}
{(\varepsilon/{\varepsilon}^{\mr{b}})}^{-\beta}$ for $\varepsilon
^\mr{b} < \varepsilon < \varepsilon ^{\mr{max}}$, 
where we set $\varepsilon ^{\mr{min}}=0.1$ eV because the synchrotron
self-absorption will be crucial below this energy \cite{Li1} and
$\varepsilon ^{\mr{max}}=1$ MeV because the pair absorption will be
crucial above this energy \cite{Asa1}. Corresponding to the observed
break energy of GRB 060218, $E^{\mr{b}}=4.9$ keV with the assumption
of the relatively low Lorentz factor, we take $\varepsilon
^{\mr{b}}=0.5$ keV in the comoving frame as a typical value throughout
the Letter. We also take $\alpha=1$ and set $\beta =2.2$ as photon
indices. Note that we may have to wait for other GRB 060218-like events to
know the reliable typical values. 

We believe not only electrons but also protons will be accelerated. 
Although the detail of acceleration mechanisms is poorly known, we
assume that the first-order Fermi acceleration mechanism works in GRBs and
the distribution of nonthermal protons is given by  
$dn_{p}/d\varepsilon _{p} \propto \varepsilon _{p}^{-2}$. 
By the condition $t_{\mr{acc}}<t_{p}$, we can estimate the
maximal energy of accelerated protons, where $t_{p}$ is the total
cooling time scale given by $t_{p}^{-1} \equiv t_{p\gamma}^{-1} + 
t_{\mr{syn}}^{-1} + t_{\mr{IC}}^{-1} + t_{\mr{ad}}^{-1}$ and the
acceleration time scale is given by $t_{\mr{acc}} = \eta \varepsilon
_{p}/eBc$. Especially, the two time scales $t_{\mr{syn}}$ 
(synchrotron cooling time) and
$t_{\mr{ad}} \approx t_{\mr{dyn}}$ 
(dynamical time)
are important in our cases. 
We can estimate the
maximum proton energy by $E_{p,\mr{max}} \approx \mr{min}[eBr/\eta,
\sqrt{6 \pi e/\sigma _{\mr{T}} B \eta} (\Gamma m_{p}^2 c^2/m_e)]$ from the 
conditions, $t_{\mr{acc}}<t_{\mr{dyn}}$ and $t_{\mr{acc}}<t_{\mr{syn}}$.
These two conditions equivalently lead to, 
\begin{eqnarray}
0.5 \eta (\Gamma/10) E_{p,20} &\lesssim& L_{\mr{M},48}^{1/2} \epsilon
  _{B,-1}^{1/2} {\left( \frac{\Gamma _{\mr{sh}}(\Gamma
  _{\mr{sh}}-1)}{2} \right)}^{1/2} \nonumber \\
  &\lesssim& 0.55 {\eta}^{-1} r_{15} {(\Gamma/10)}^3 E_{p,20}^{-2}, \label{pro}
\end{eqnarray}
where we have used notations such as 
$E_{p} \equiv {10}^{20} \, \mr{eV} E_{p,20}$. 
These inequalities suggest that the only relatively more
luminous/magnetized LL-GRBs with higher Lorentz factor 
(i.e., larger $L_{\rm M}$ and/or $\epsilon_B$, and higher $\Gamma$)
will have possibilities to explain the observed flux of UHECRs.

We consider 
neutrinos from the decay of pions generated by photomeson productions.
The photomeson time scale is $t_{p\gamma}$.
Let us evaluate $f_{p\gamma} \equiv t_{\mr{dyn}}/t_{p\gamma}$ analytically
using the $\Delta$-resonance approximation \cite{KM2,Wax1} as,
\begin{equation}
f_{p\gamma} \simeq 0.06 \frac{L_{\mr{max},47}}{r_{15} {(\Gamma/10)}^2
E_{5 \, \mr{keV}}^{\mr{b}}} \left\{ \begin{array}{rl} 
{(E_p/E_p^\mr{b})}^{\beta-1} & \mbox{($E_p < E_{p}^{\mr{b}}$)}\\
{(E_{p}/E_{p}^{\mr{b}})}^{\alpha-1} & \mbox{($E _p^{\mr{b}} < E_p$)} 
\end{array} \right. \label{pgamma1}
\end{equation}
where $E_{p}^{\mr{b}} \simeq 0.5 \bar{\varepsilon}
_{\Delta}m_pc^2 \Gamma ^2/E^{\mr{b}}$ is the
proton break energy. Here, $\bar{\varepsilon}
_{\Delta}$ is around $0.3$ GeV. From Eq. (\ref{pgamma1}), we can
conclude that a 
moderate fraction of high energy accelerated protons will be converted
into neutrinos.

Next, we consider the contribution to the neutrino flux from a thermal
photon component. The discovery of
the thermal component in GRB 060218 will provide additional photon flows. This
photon flow has a possibility to produce more neutrinos by
interaction with protons accelerated in internal shocks.
We take $kT=0.15$ keV and $r_{\mr{BB}}={10}^{12}$ cm as the
typical photon energy of the thermal component and the apparent emitting
radius \cite{Cam1}, respectively. Just for simplicity, we assume the photon
density drops as $\propto r^{-2}$ and approximate it by the
isotropic distribution with $dn/d\varepsilon (\varepsilon) \approx
dn_{\mr{lab}}/d\varepsilon _{\mr{lab}}(\varepsilon _{\mr{lab}})$, 
where $dn_{\mr{lab}}/d\varepsilon _{\mr{lab}}$ is the photon distribution in
the laboratory frame. 

\section{Results and Discussions}
We calculate neutrino spectra for some parameter sets and will show
the case where the width of shells $\Delta \approx (r/2\Gamma ^2) =
4.5 \times {10}^{12}$ cm, according to $\delta t \sim 150$ s. In our calculations, we include various cooling processes of pions 
(synchrotron cooling, IC cooling, adiabatic cooling) similarly to
Murase \& Nagataki (2006a,2006b). These cooling processes are important
for neutrino spectra \cite{Rac1,Wax1}.
A diffuse neutrino background under the standard $\Lambda$CDM cosmology
$(\Omega _{\mr{m}}=0.3, \Omega _{\Lambda}=0.7; H_{0}=71 \, \mr{km
\, s^{-1} \, Mpc^{-1}})$ is calculated by using Eq. (15) of Murase \&
Nagataki (2006a), where we set $z_{\mr{max}}=11$. 
Assuming that the long GRB rate traces the starformation rate (SFR),
we exploit the SF2 model of Porciani \& Madau (2001) combined with the 
normalization of geometrically corrected overall HL-GRB rates 
$R_{\mr{HL}}(0)$ obtained by Guetta et
al. (2005) for HL-GRBs. The local LL-GRB rate is very uncertain for
now. Soderberg et al. (2006b) obtained the geometrically corrected overall GRB
rate, $R_{\mr{LL}}(0)=230 \,
\mr{Gpc}^{-3} \mr{yr}^{-1}$. Liang et al. (2006a) also had a
high value, $\rho _{\mr{LL}}(0) = 550 \, \mr{Gpc}^{-3} \mr{yr}^{-1}$. 
(Note that the true rate $R_{\mr{LL}}(0)$ is almost the same as the
apparent one $\rho_{\mr{LL}}(0)$ for LL-GRBs because we are assuming
GRB 060218-like spherical bursts.)
However, too large rates will be impossible due to 
constraints by observations of SNe \Roman{ichi}bc. Soderberg et
al. (2006a) argued that at most $\sim 10 \, \%$ of SNe \Roman{ichi}bc are
associated with off-beam LL-GRBs based on their late-time radio
observations of 68 local SNe \Roman{ichi}bc. Hence, the most
optimistic value allowed from the local SNe \Roman{ichi}bc rate 
will be around $\sim 4800 \, \mr{Gpc}^{-3} \mr{yr}^{-1}$ (and the
larger value is ruled out with a confidence level of $\sim 90 \, \%$)
\cite{
Sod2}. 
The high rate might be realized if 
LL-GRBs are related with the birth of magnetars
and the fraction of SNe \Roman{ichi}bc that produce magentars 
is comparable with that of SNe \Roman{ni}, i.e., $\sim 10\%$.
\begin{figure}[t]
\includegraphics[width=\linewidth]{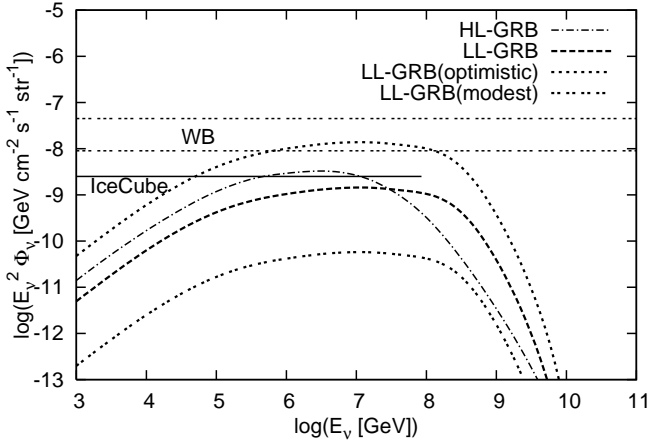}
\caption{\label{Fig1} The neutrino background from
GRBs for $\xi _{\mr{acc}}=10$ and $\xi _{B}=1$. LL-GRB: $r= 9 \times 
{10}^{14} \, \mr{cm}$ and $\Gamma =10$ with the local rate 
$\sim 500 \, \mr{Gpc}^{-3} \mr{yr}^{-1}$ obtained by
Liang et al. (2006a). LL-GRB (optimistic): $r= 9 \times
{10}^{14} \, \mr{cm}$ and $\Gamma=10$ with the most optimistic local
rate $\sim 4800 \, \mr{Gpc}^{-3} \mr{yr}^{-1}$. LL-GRB (modest): $r= 9 \times
{10}^{14} \, \mr{cm}$ and $\Gamma=10$ with the modest local
rate $\sim 20 \, \mr{Gpc}^{-3} \mr{yr}^{-1}$. 
HL-GRB: taken from \cite{KM1} with
$E_{\gamma,\mr{iso}}/N=2 \times
{10}^{51} \, \mr{ergs}$, $r= {10}^{13-14.5} \, \mr{cm}$ and
$\Gamma=300$. WB: Waxman-Bahcall bounds \cite{Wax2}. $\xi _{B}$ and 
$\xi_{\mr{acc}}$ are the ratio of energy density,
$\xi _{B} \equiv U_{B}/U_{\gamma}$ and
$\xi _{\mr{acc}} \equiv U_{p}/U_{\gamma}$, respectively. 
For the fast cooling case and the acceleration efficiency $\sim 1$, we 
have $\xi _{B} \sim (\epsilon _{B}/\epsilon _{e})$ and $\xi
_{\mr{acc}} \sim 1/\epsilon _{e}$.}
\end{figure}

Although we calculate numerically, we can estimate the diffuse neutrino
flux from LL-GRBs approximately by the following analytical expression 
\cite{KM2,Wax2},
\begin{eqnarray}
E_{\nu}^2\Phi _{\nu} &\sim& \frac{c}{4\pi H_{0}}
 \frac{1}{4} \mr{min}[1,f_{p\gamma}] E_{p}^2 \frac{dN_{p}}{dE_{p}}
 R_{\mr{LL}}(0) f_{z} \nonumber\\
&\simeq& 7 \times 10^{-10} \mr{GeV cm^{-2} s^{-1} str^{-1}} \, 
\left( \frac{\xi _{\mr{acc}}}{10} \right) E_{\mr{LL},50} \nonumber\\
&\times& \left(\frac{f_{p\gamma}}{0.05}\right) \left(
\frac{R_{\mr{LL}}(0)}{500 \,
 \mr{Gpc}^{-3}\mr{yr}^{-1}}\right)\left( \frac{f_{z}}{3} \right),
\end{eqnarray}
where $E_{\mr{LL}}$ is the geometrically corrected radiated energy of
LL-GRBs, $f_{z}$ is the correction factor for the possible contribution from
high redshift sources, and we have used $\varepsilon _{p,\mr{max}} 
\sim {10}^{9}$ GeV.
Our numerical results are shown in Fig. 1. From these results, 
we can estimate the number of muon events $N_{\mu}$ 
due to muon-neutrinos above TeV energy by using Eq. (18) of Ioka et
al. (2005) as the detection probability and a geometrical detector area of 
$A_{\mr{det}}=1 {\mr{km}}^{2}$. 
From Fig. 1, we can obtain $N_{\mu}= 6.6$ events/yr for $\rho
_{\mr{LL}}(0)=500 \, {\mr{Gpc}}^{-3} {\mr{yr}}^{-1}$ 
and $N_{\mu}=64$ events/yr for the most optimistic local rate. We also
show the modest case where the local rate of LL-GRBs is comparable to
the geometrically corrected local rate of HL-GRBs, $R_{\mr{HL}}(0)
\sim 20 \, \mr{Gpc}^{-3} \mr{yr}^{-1}$. In this case, we can find 
$N_{\mu}=0.3$ events/yr. The neutrino backgrounds from LL-GRBs have
possibilities to be comparable with that from HL-GRBs, 
$N_{\mu}=17$ events/yr \cite{KM1}.
\begin{figure}[b]
\includegraphics[width=\linewidth]{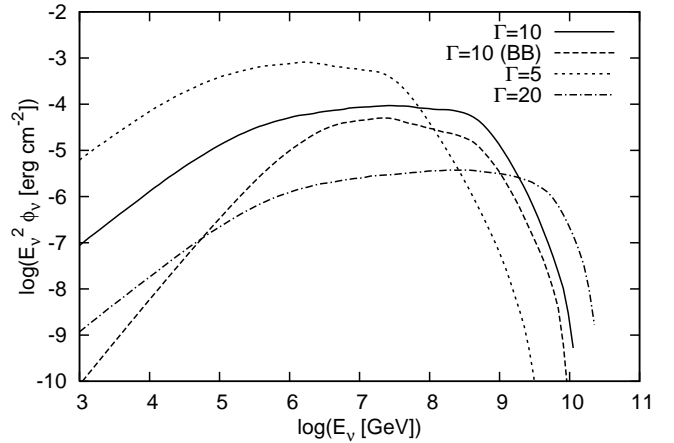}
\caption{\label{Fig2} The observed muon-neutrino $(\nu _{\mu} +
\bar{\nu} _{\mu})$ spectra for one very nearby GRB event at $10$
Mpc. Solid line: $r = 9 \times {10}^{14} \, \mr{cm}$ and $\Gamma
=10$. Dashed line: the
contribution from the blackbody target photon 
with $r = 9 \times {10}^{14} \,
\mr{cm}$ and $\Gamma=10$. Dotted line: $r = 2.25 \times {10}^{14} \,
\mr{cm}$ and $\Gamma=5$. Dot-dashed line: $r = 3.6 \times {10}^{15} \,
\mr{cm}$ and $\Gamma=20$. In all cases $\xi _{B}=1$ and
$\xi_{\mr{acc}}=10$ (see the caption of Fig.1). Note, the case
of $\Gamma=20$ would not be plausible because the magnetic
field strength seems too small to explain the prompt
emission by the standard model.}
\end{figure}

Unfortunately, neutrino signals from LL-GRBs are dark in the sense 
that most signals will not
correlate with the prompt emission. Only for very nearby bursts, we
might be able to expect their
correlations and it will need many-years operations. The BAT detector on 
\textit{Swift} has the sensitivity to detect the
bursts $\gtrsim {10}^{-8} \, \mr{ergs} \, \mr{cm}^{-2} \,
\mr{s}^{-1}$. Hence, we can expect correlated
events only when $d_{\mr{L}} \lesssim 300$ Mpc for bursts with
$L_{\mr{max}} \sim {10}^{47}$ ergs/s. The expected correlated muon
events are $N_{\mu} \sim 1$ events per $11$ years for 
$\rho _{\mr{LL}}(0)=500 \, {\mr{Gpc}}^{-3} {\mr{yr}}^{-1}$
. 

However, SNe \Roman{ichi}bc associated with LL-GRBs could be detected
by optical-infrared follow-ups triggered by a neutrino event.
The angular resolution of IceCube for neutrinos
is about $1$ degree or so, which can be searched with wide-field cameras such
as Suprime-Cam on the Subaru telescope (whose field-of-view is 0.5 degrees) up
to $z \sim 1.2$. 
In the field-of-view we would find $\sim 10$ SNe
and $\sim 1$ SNe \Roman{ichi}bc that exploded within $\sim 1$ month. 
With the SN light curves $\sim 10$ days 
after the burst, we can pin down the burst time within $\sim 1$ day or so, during which the atmospheric neutrino background within 
$1$ degree would be small, i.e., $\lesssim 0.1$ events/day for above TeV energy neutrinos and less for higher energy threshold
\cite{And1}. In addition, SNe \Roman{ichi}bc could be specified by 
using telescopes such as HST. Therefore, we can in principle detect LL-GRB neutrino events associated with SNe \Roman{ichi}bc, 
even though X/$\gamma$-rays are not observed by \textit{Swift}. The 
expected number of
muon events is  $N_{\mu}=2.4$ events/yr for LL-GRBs within $z \sim
1.2$, with $r= 9 \times {10}^{14}$ cm, $\Gamma=10$, and 
$\rho _{\mr{LL}}(0)=500 \, {\mr{Gpc}}^{-3} {\mr{yr}}^{-1}$. 
Of course, such a follow-up with SNe detections will be difficult
and it is severer to distinguish SNe \Roman{ichi}bc from SNe
\Roman{ichi}a at higher redshift. Nevertheless, it is worthwhile to
develop this kind of possibility of high energy neutrino
astronomy not only for finding far SNe Ibc associated with LL-GRBs but also for revealing their origins. 

We can expect high energy neutrinos from 
one LL-GRB only if the burst is nearby or energetic, similarly to the
case of HL-GRBs. In Fig. 2, we show an
example of the observed neutrino spectra from the source at $10$ Mpc. The 
expected muon events from neutrinos above TeV energy are $N_{\mu}=1.1$ 
events in the case of $\Gamma=10$ in Fig. 2. If we can detect such an
event, we will be able to obtain some information on 
$\xi_{\mr{acc}}$, $\xi _{B}$, the photon density, the duration of
bursts, and so on. 
In Fig. 2, we also show the contribution from the thermal target photon. The
GRB 060218-like bursts could provide us $N_{\mu}=0.2$
events originating from the interaction between nonthermal protons
and the thermal photon flow. This result depends on the temperature of 
the black body region. 

HL-GRBs may be the main sources of UHECRs \cite{Wax4}. The optical 
thickness for the photomeson production can be smaller than unity 
especially at larger radii $r \gtrsim {10}^{14}{(E_{\gamma,\mr{iso}}
/N{10}^{51} \, \mr{ergs})}^{1/2} \, \mr{cm}$ in the internal shocks of
HL-GRBs and UHECRs can be produced in such regions. In the case of
LL-GRBs, it seems more difficult
to accelerate protons up to ultra-high energy due to the lowness of
their luminosities and it would need some fine tuning (see
Eq. \ref{pro}), although we have to know about their
properties such as their luminosity function. Even if the acceleration
to $\sim {10}^{20}$ eV is difficult, the energy budget of LL-GRBs
could be large enough to explain UHECRs ($\sim {10}^{44} \, \mr{ergs}
\, \mr{Mpc}^{-3} \, \mr{yr}^{-1}$) because the cosmic-ray
production rate per $\mr{Mpc}^{3}$ is estimated by,
\begin{eqnarray}
E_{p}^2\frac{d\dot{N}_p}{dE_{p}^2} \sim &2.5& \times {10}^{43} \, \mr{ergs}
\, \mr{Mpc}^{-3} \, \mr{yr}^{-1} {\left(\frac{\xi _{\mr{acc}}}{10}
\right)} \nonumber \\
&\times& N L_{\mr{max},47} r_{15} {\left( \frac{\Gamma}{10}
\right)}^{-2} \left( \frac{{\rho}_{\mr{LL}}(0)}{500 \,  \mr{Gpc}^{-3}
\, \mr{yr}^{-1}} \right).
\end{eqnarray}
Therefore, when the maximum proton energy exceeds ${10}^{18.5}$ eV, the
neutrino flux should be constrained by the observed flux of UHECRs,
even if LL-GRBs cannot explain all UHECRs. This implies that we have
possibilities to constrain physical parameters of LL-GRBs by the
observation of UHECRs.

High energy neutrino emission cannot avoid high energy gamma-ray
emission through the neutral pion decay. Such high gamma-rays would
cascade in the source and/or in microwave and infrared
background \cite{Der1,Raz3}. The detailed calculation is needed to
calculate the expected spectra. However, the detection of such high energy 
emission by GLAST and/or the BAT detector on \textit{Swift} would be
difficult except for nearby and/or energetic bursts similarly to
the cases of neutrinos.

If there is a shock break out which might be the origin of the
mysterious thermal component, the shock may become collisionless,
and protons may be accelerated there as well as electrons, so that neutrinos
could be produced through the $pp$ interaction \cite{Wax3}. In addition,
 protons accelerated by internal shocks inside the stellar envelope
could produce detectable $\sim $ TeV neutrinos mainly by $pp$ interactions 
\cite{Mes1,Raz2,And1}. These other neutrino signals might become clues on
the connection between GRBs and SNe.  

In this Letter, we have discussed a possibility
that LL-GRBs could produce UHECRs and detectable high energy neutrinos.
Because the possible higher rate of LL-GRBs can
cover the relatively lower energy of them, we can expect 
that the diffuse neutrino flux could be comparable with that of HL-GRBs.
Of course, the results depend on several unknown parameters such as the bulk
Lorentz factor, and future observations are needed for more realistic 
predictions. If parameters we have adopted are typical, expected
neutrino signals may
not only give us independent information but could also be useful as one of
indicators of far SNe \Roman{ichi}bc and may contribute to revealing
their mysterious origin. One of possibilities is the birth of
magnetars \cite{Sod1}. Possibly, metallicity might play a
crucial role 
\cite{Sta1}.
     
We thank the referee, M. Doi, N. Tominaga, R. Romani, S. Jha, K. Toma,
N. Kawanaka, J.F. Beacom, and E. Liang for helpful advices. 
This work is supported by Grants-in-Aid for Scientific Research
of the Japanese Ministry of Education, Culture, Sports, Science,
and Technology 18740147 (K.I.).

\clearpage





\end{document}